\documentstyle[12pt]{article}

\newcommand {\nn}    {\nonumber}
\newcommand {\vs}[1]  { \vspace*{#1 cm} }

\newcounter{eq}
\newcounter{sc}

\newcommand {\PL}   {Phys.Lett.}
\newcommand {\PR}   {Phys.Rev.}
\newcommand {\PRL}   {Phys.Rev.Lett.}

\def\overleftrightarrow#1{\vbox{\ialign{##\crcr
 $\leftrightarrow$\crcr\noalign{\kern-1pt\nointerlineskip}
 $\hfil\displaystyle{#1}\hfil$\crcr}}}

\newlength{\minitwocolumn}
\begin{document}

\begin{flushright}
EDO-EP-35\\
December, 2000\\
\end{flushright}
\vspace{30pt}

\pagestyle{empty}
\baselineskip15pt

\begin{center}
{\large\bf Localization of Bulk Fields on $AdS_4$ Brane
in $AdS_5$

 \vskip 1mm
}

\vspace{20mm}

Ichiro Oda
          \footnote{
          E-mail address:\ ioda@edogawa-u.ac.jp
                  }
\\
\vspace{10mm}
          Edogawa University,
          474 Komaki, Nagareyama City, Chiba 270-0198, JAPAN \\

\end{center}


\vspace{15mm}
\begin{abstract}
We study the localization of various bulk fields on an $AdS_4$ brane
in $AdS_5$. In this case, for a small brane cosmological constant,
all the bulk fields ranging from scalar to graviton are naturally
confined to the brane only through the gravitational interaction.
In particular, for the vector field we can find an interesting
zero mode solution which satisfies the box boundary condition.
In the cases of spin 1/2 spinor and 3/2 gravitino fields, the 
form of zero modes is very similar to as in a flat Minkowski brane,
but they are trapped on the $AdS_4$ brane even without introducing
a mass term with a 'kink' profile.

\vspace{15mm}

\end{abstract}

\newpage
\pagestyle{plain}
\pagenumbering{arabic}

\rm

The past few years have witnessed a lot of interest in the study
of a brane world where the standard model is assumed to live
on a brane embedded in a higher dimensional space-time whereas 
the gravitational field is free to propagate in the whole space-time.
Contact with observations obviously requires that the four dimensional
gravitational theory is not only reproduced on the brane but also
the standard model particles are confined on the brane by some ingenious
mechanism. The most interesting possibility is provided by superstring
theory or M-theory, but it is at present far from complete to understand 
the precise procedure of deriving a desired model of the brane world 
from superstring theory even if there have been many efforts and 
attempts so far. Then it is a natural step to pursue a different 
possibility where the local field theory gives us such a mechanism. 

Indeed, Randall and Sundrum have found a solution to the five dimensional 
Einstein's equations with a Minkowski flat 3-brane in $AdS_5$ and shown that
the effects of the four dimensional gravity on the brane is reproduced
without the need to compactify the fifth dimension \cite{Randall1, Randall2}.
(This model is generalized to the case of many branes in 
Ref. \cite{Oda1, Oda2}.)  Regarding the localization of the standard
model particles, it has been also shown that the local field theory can
present
a nice mechanism in an arbitrary space-time dimension \cite{Oda3}.

Recently, Karch and Randall have found a unexpected result that four
dimensional 
gravity is induced even on an $AdS_4$ brane in an $AdS_5$ background owing to 
the locaization of a massive and normalizable bound state \cite{Karch}. 
(See also related works \cite{Kogan, Porrati, Miemiec, Schwartz, Grassi}.) 
In this model, in order to avoid the conflict with experiment, 
a four dimensional cosmological constant is taken to be small 
in Planck units, so the massive bound state is effectively massless 
at large scales. Then, a question, which naturally arises, is whether or not 
the standard model particles are also localized on the $AdS_4$ brane 
in terms of the gravitational interaction as in a Minkowski 3-brane.
This is precisely the question with which we are concerned in this article.
We will see below that the whole particles are trapped on the brane in a
remarkable way.  

Let us first fix our physical setup. Following the work of Karch and Randall
we shall consider a five dimensional anti-de Sitter space-time ($AdS_5$)
with a warp factor
\begin{eqnarray}
ds^2 = e^{2A(z)} \left(\hat{g}_{\mu\nu}(x^\lambda) dx^\mu dx^\nu 
+ dz^2 \right),
\label{1}
\end{eqnarray}
where $\mu, \nu, \lambda = 0, 1, 2, 3$. Here the metric $\hat{g}_{\mu\nu}$
on a 3-brane denotes an $AdS_4$ background. Moreover, the warp 
factor $A(z)$ in the conformal coordinate $z$ is given by
\begin{eqnarray}
e^{A(z)} = \frac{L \sqrt{-\Lambda}}{\sin \sqrt{-\Lambda}(|z| + z_0)},
\label{2}
\end{eqnarray}
with the five dimensional cosmological constant $\Lambda_5$
being connected with a constant $L$ by $\Lambda_5 = - \frac{3}{L^2}$.
Now the four dimensional cosmological constant $\Lambda$
is assumed to be small and negative. In this article, we follow the 
standard conventions and notations of the textbook of 
Misner, Thorne and Wheeler \cite{Misner}. 
Also note that a single positive tension brane sits at the origin $z=0$ and 
orbifold boundary conditions are imposed in a usual way. From the form
of $e^{A(z)}$ and orbifold boundary conditions, we can limit ourselves
to the region $0 < z < \frac{\pi}{\sqrt{-\Lambda}} - z_0$. Instead of the $z$
coordinate, we sometimes make use of the coordinate $w$ whose definition
is given by $w \equiv \sqrt{-\Lambda} z$ (and $\varepsilon \equiv 
\sqrt{-\Lambda} z_0$), so in the $w$ coordinate $0 < w < \pi - \varepsilon$. 
Since we have fixed our physical setup, in what follows, we shall study 
the property of localization of bulk fields according to their spin in order. 
As a remark, in this article, we will assume that the background
metric is not modified by the presence of the bulk fields, namely, we will
neglect the back-reaction on the metric from the bulk fields. 

We are now ready to start with the case of a massless real scalar field of
spin 0. 
The extension to complex and/or massive scalar fields is straightforward. 
The action takes the form
\begin{eqnarray}
S_0 = -\frac{1}{2} \int d^5 x \sqrt{-g} g^{MN} \partial_M \Phi
\partial_N \Phi,
\label{3}
\end{eqnarray}
where $M, N$ denote the five dimensional space-time indices. From 
this action, we have the equation of motion, 
$\frac{1}{\sqrt{-g}} \partial_M(\sqrt{-g}g^{MN}\partial_N \Phi) = 0$.
With the metric ansatz (\ref{1}), this equation of motion leads to
\begin{eqnarray}
\left[ - \partial_z^2 + \frac{9}{4} A'(z)^2 + \frac{3}{2} A''(z)
\right] \tilde{\Phi} = m^2 \tilde{\Phi},
\label{4}
\end{eqnarray}
where we have defined $\tilde{\Phi}$ as $\tilde{\Phi} \equiv 
e^{\frac{3}{2} A(z)} \Phi$ and assumed that $\Box \tilde{\Phi}
\equiv m^2 \tilde{\Phi}$ with the definition of 4D d'Alembertian
operator $\Box \equiv \frac{1}{\sqrt{-\hat{g}}} \partial_\mu
(\sqrt{-\hat{g}} \hat{g}^{\mu\nu}\partial_\nu)$. The prime denotes
differentiation with respect to the coordinate $z$. Eq. (\ref{4})
can be cast to the Schr\"{o}dinger-like equation in the $w$ coordinate
with help of Eq. (\ref{2}) as follows:
\begin{eqnarray}
\left[ - \partial_w^2 + V(w) \right] \tilde{\Phi} = E \tilde{\Phi},
\label{5}
\end{eqnarray}
where $E \equiv |\Lambda| m^2$ and the potential $V(w)$ is of the 
form
\begin{eqnarray}
V(w) =  - \frac{9}{4} + \frac{15}{4} \frac{1}{\sin^2(|w|+ \varepsilon)}
- 3 \cot(\varepsilon) \delta(w).
\label{6}
\end{eqnarray}
In the case where $\varepsilon = \frac{\pi}{2}$, the delta function
decouples, thereby implying that there is no brane, so we can find a 
general solution in terms of the hypergeometric function
\begin{eqnarray}
\tilde{\Phi}(w) = A_1 \frac{1}{(\sin w)^{\frac{3}{2}}} F(- \frac{3}{4}
+ \frac{\sqrt{4E+9}}{4}, - \frac{3}{4} - \frac{\sqrt{4E+9}}{4}, \frac{1}{2},
\cos^2 w) \nn\\
+ A_2 \frac{\cos w}{(\sin w)^{\frac{3}{2}}} F(- \frac{1}{4}
+ \frac{\sqrt{4E+9}}{4}, - \frac{1}{4} - \frac{\sqrt{4E+9}}{4}, \frac{3}{2},
\cos^2 w),
\label{7}
\end{eqnarray}
where $A_1, A_2$ are integration constants. To gain a solution to Eq. 
(\ref{6}), one needs to impose the boundary conditions at $w = 0$ and 
$w = \pi - \varepsilon$, but it turns out that it is a delicate problem
and consequently only the numerical analysis is available \cite{Karch}.

In this respect, we make use of an exactly solvable toy model sharing
the similar qualitative features with the original model 
(\ref{5}) \cite{Schwartz}.
In the toy model, the potential takes the form $V(w) = - 3 \cot(\varepsilon) 
\delta(w)$ and contains a wall at $w = \pi - \varepsilon$ \cite{Schwartz}.
Then there is a bound state solution $\chi_0(w)$ with energy $E = -\kappa^2$
\begin{eqnarray}
\chi_0(w) = \sinh \kappa (\pi - \varepsilon - w)
\label{8}
\end{eqnarray}
where $\kappa$ must satisfy the equation
\begin{eqnarray}
\frac{2}{3} \tan \varepsilon = \frac{1}{\kappa} \tanh \kappa (\pi 
- \varepsilon)
\label{9}
\end{eqnarray}
In addition to this bound state, it is easy to find a set of massive
modes with energy $E = k^2$, which are given by
\begin{eqnarray}
\chi_k(w) = \sin k (\pi - \varepsilon - w)
\label{10}
\end{eqnarray}
where $k$ must satisfy the equation
\begin{eqnarray}
\frac{2}{3} \tan \varepsilon = \frac{1}{k} \tan k (\pi 
- \varepsilon)
\label{11}
\end{eqnarray}
These results are similar to those of the gravitational field 
\cite{Karch, Schwartz}. Actually it is known that transverse traceless 
graviton modes in general obey the equation of a massless scalar in 
a curved background, so the results obtained here merely confirms this
fact in an explicit manner. But owing to the existence of the brane
cosmological constant there is a slight differece between
the two cases, which is that in the case at hand $\Box \tilde{\Phi}
= m^2 \tilde{\Phi}$ while in the gravity case $(\Box + 2 |\Lambda|)
h^{TT}_{\mu\nu} = m^2 h^{TT}_{\mu\nu}$.

Let us turn our attention to the localization of a scalar
field on an $AdS_4$ brane in $AdS_5$. In the situation at hand,
let us focus on only the bound state (\ref{8}) because this bound state
corresponds to the massive graviton mode and the treatment of
a set of massive modes can be done in an analogous way.
{}Following the method \cite{Bajc, Oda4}, let us plug the bound state 
solution $\Phi_0(x^M) = \phi(x^\mu) \chi_0(z) e^{-\frac{3}{2}A(z)}$
into the starting action:
\begin{eqnarray}
S_0^{(0)} &=& -\frac{1}{2} \int d^5 x \sqrt{-g} g^{MN} \partial_M \Phi_0
\partial_N \Phi_0 \nn\\
&=& -\frac{1}{2} \int d^4 x \sqrt{-\hat{g}} \hat{g}^{\mu\nu} 
\partial_\mu \phi \partial_\nu \phi \int_0^{\frac{\pi}{\sqrt{-\Lambda}}-z_0}
dz \chi_0^2(z).
\label{12}
\end{eqnarray}
Here it is worthwhile to notice that the condition that the bound state  
solution is localized on a brane is equivalent to the normalizability 
of the state wave function on the brane \cite{Bajc, Oda4}. This condition
now amount to the finiteness of an integral over $z$ coordinate in Eq.
(\ref{12}).
Indeed, we can show that the integral is finite as follows:
\begin{eqnarray}
I_0 &{}& \equiv \int_0^{\frac{\pi}{\sqrt{-\Lambda}}-z_0} dz \chi_0^2(z)
\nn\\
&{}& = \frac{1}{4 \kappa \sqrt{- \Lambda}} \left[\sinh 2 \kappa (\pi -
\varepsilon) - 2 \kappa (\pi - \varepsilon) \right] < \infty.
\label{13}
\end{eqnarray}
Note that $I_0$ is strictly finite as long as the brane cosmological
constant $\Lambda$ is nonzero.
Thus the bound state of scalar field is confined on an $AdS_4$ brane
in $AdS_5$ by the gravitational interaction. This proof obviously means that
transverse, traceless, massive modes of the gravitational field
are also confined to the brane.   

Next let us turn to spin 1/2 spinor field. The starting action
is the conventional Dirac action with a mass term:
\begin{eqnarray}
S_{1/2} = \int d^5 x \sqrt{-g} \bar{\Psi} i (\Gamma^M D_M
+ m \varepsilon(z)) \Psi, 
\label{14}
\end{eqnarray}
where the covariant derivative is defined as $D_M \Psi = ( \partial_M + 
\frac{1}{4} \omega_M^{AB} \gamma_{AB} ) \Psi$ with the definition of
$\gamma_{AB} = \frac{1}{2} [\gamma_A, \gamma_B]$, and $\varepsilon(z)$
is $\varepsilon(z) \equiv \frac{z}{|z|}$ and $\varepsilon(0) \equiv 0$.
Here the indices $A, B$ are the ones of the local Lorentz frame and 
the gamma matrices $\Gamma^M$ and $\gamma^A$ are related by the 
vielbeins $e_A^M$ through the usual relations $\Gamma^M = e_A^M \gamma^A$ 
where  $\{\Gamma^M, \Gamma^N\} = 2 g^{MN}$ and $\{\gamma^A, \gamma^B\} 
= 2 \eta^{AB}$.
A feature of the action is the existence of a mass term with 
a 'kink' profile. We have just introduced this type of mass term in the 
action since the existence has played a critical role 
in the localization of fermionic
fields on a Minkowski brane in an arbitrary dimension \cite{Jackiw, Grossman, 
Oda3}.
However, as will be shown in what follows, in the case of an $AdS$ brane 
such the mass term does not play any important role. This is one of
interesting
things in the model at hand compared to a flat Minkowski brane.

In the background (\ref{1}), the torsion-free conditions yield an explicit
expression of the spin connections:
\begin{eqnarray}
\omega_\mu = \frac{1}{2} A'(z) \gamma_\mu \gamma_z + 
\hat{\omega}_\mu(\hat{e}), \ \omega_z = 0,
\label{15}
\end{eqnarray}
where we have defined $\omega_M \equiv \frac{1}{4} \omega_M^{AB} \gamma_{AB}$,
$\hat{\omega}_\mu(\hat{e}) \equiv \frac{1}{4} \hat{\omega}_\mu^{ab}(\hat{e}) 
\gamma_{ab}$, and $\gamma_\mu \equiv \hat{e}_\mu^a \gamma_a$.
Using Eq. (\ref{15}), the Dirac equation $(\Gamma^M D_M + m \varepsilon(r)) 
\Psi = 0$ stemming from the action (\ref{14}) can be cast to the form
\begin{eqnarray}
\left[\Gamma^z (\partial_z + 2 A')  + m \varepsilon(z) 
+ \Gamma^\mu (\partial_\mu + \hat{\omega}_\mu)  \right] \Psi = 0. 
\label{16}
\end{eqnarray}
Let us find the massless zero-mode solution with the form of $\Psi(x^M) = 
\psi(x^\mu) u(z)$ such that $\Gamma^\mu \hat{D}_\mu \psi \equiv \Gamma^\mu
(\partial_\mu + \hat{\omega}_\mu) \psi = 0$ and the chirality condition 
$\Gamma^z \psi = \psi$ is imposed on the brane fermion. Then, Eq. (\ref{16})
is reduced to a first-order differential equation to $u(z)$ and is 
easily integrated to be 
\begin{eqnarray}
u(z) = u_0 e^{-2 A(z) - m \varepsilon(z) z},
\label{17}
\end{eqnarray}
with an integration constant $u_0$. 

In order to check the localization of this mode, let us
insert this solution into the Dirac action (\ref{14}). Then the action 
reduces to the form
\begin{eqnarray}
S_{1/2}^{(0)} &=& \int d^5 x \sqrt{-g} \bar{\Psi}^{(0)} i (\Gamma^M D_M
+ m \varepsilon(z)) \Psi^{(0)} \nn\\
&=& \int d^4 x \sqrt{-\hat{g}} \bar{\psi} i \gamma^\mu \hat{D}_\mu \psi
\int_0^{\frac{\pi}{\sqrt{-\Lambda}}-z_0} dz e^{4A(z)} u^\dagger(z) u(z)
+ \cdots. 
\label{18}
\end{eqnarray}
Again the condition of the trapping of the bulk spinor on an $AdS_4$ brane 
requires that an integral over $z$ has a finite value. The integral
is easily evaluated as follows:
\begin{eqnarray}
I_{1/2} &{}& \equiv \int_0^{\frac{\pi}{\sqrt{-\Lambda}}-z_0} dz 
e^{4A(z)} u^\dagger(z) u(z) \nn\\
&{}& = |u_0|^2 \int_0^{\frac{\pi}{\sqrt{-\Lambda}}-z_0} dz 
e^{- 2 m z} < \infty.
\label{19}
\end{eqnarray}
Here one interesting thing has happened. Recall that in a Minkowski
brane, only the massive bulk fermion with a 'kink' profile is 
localized on the brane whereas the massless one is not so. 
This fact can be traced in Eq. (\ref{19}) since $I_{1/2}$ at $\Lambda = 0$
is divergent in the massless limit $m=0$.
(Note that in both the Minkowski brane and the AdS brane, the form of 
the zero-mode solution of fermion, (\ref{17}), is common so this 
consideration is legitimate.) In the case at hand, irrespective
of the presence of mass term, the bulk spinor can be localized on
the brane through the gravitational interaction as long as the
brane cosmological constant is nonvanishing.

Now we are willing to consider spin 1 $U(1)$ vector field. Incidentally
the generalization to the nonabelian gauge fields is straightforward. 
And the inclusion of bulk mass does not change the results obtained below.
The action reads
\begin{eqnarray}
S_1 = -\frac{1}{4} \int d^5 x \sqrt{-g} g^{MN} g^{RS} F_{MR} F_{NS}, 
\label{20}
\end{eqnarray}
where $F_{MN} = \partial_M A_N - \partial_N A_M$. The equations of motion
$\partial_M (\sqrt{-g} g^{MN} g^{RS} F_{NR}) = 0$ can be solved under
the gauge condition $A_z = 0$. The solution with the form $A_\mu(x^M) = 
a_\mu(x^\lambda) \rho(z)$ is searched where $a_\mu$ satisfy the
equations $\hat{\nabla}^\mu a_\mu = \partial^\mu f_{\mu\nu} = 0$ with
the definition of $f_{\mu\nu} \equiv \partial_\mu a_\nu - \partial_\nu a_\mu$.
Then, the Maxwell equations are reduced to a single differential
equation:
\begin{eqnarray}
\partial_z ( e^{A(z)} \partial_z \rho(z) ) = 0. 
\label{21}
\end{eqnarray}
In the case of a Minkowski brane, we have selected a constant
zero-mode solution $\rho(z) = const$, which leads to non-localization
of the vector field. On the other hand, in an $AdS$ brane a new solution
is available, which is given by $e^{A(z)} \partial_z \rho(z) = const
\not= 0$.
(Note that this solution is not localized on a Minkowski brane,
either.)  As a result, we obtain a solution to Eq. (\ref{21}):
\begin{eqnarray}
\rho(z)  = - \frac{\alpha}{L (- \Lambda)} \cos \sqrt{- \Lambda}
(z + z_0) + \beta, 
\label{22}
\end{eqnarray}
where $\alpha, \beta$ are integration constants.

Let us investigate whether this solution is localized on an $AdS_4$
brane or not according to the method used above. 
The substitution of this solution into the action leads to
\begin{eqnarray}
S_1^{(0)} &=& -\frac{1}{4} \int d^5 x \sqrt{-g} g^{MN} g^{RS} F_{MR}^{(0)}
F_{NS}^{(0)} \nn\\ 
&=& -\frac{1}{4} \int d^4 x \sqrt{-\hat{g}} \hat{g}^{\mu\nu} 
\hat{g}^{\lambda\sigma} f_{\mu\lambda}f_{\nu\sigma}
\int_0^{\frac{\pi}{\sqrt{-\Lambda}}-z_0} dz e^{A(z)} \rho^2(z) \nn\\
&{}& -\frac{1}{4} \int d^4 x \sqrt{-\hat{g}} \hat{g}^{\mu\nu} 
a_\mu a_\nu \int_0^{\frac{\pi}{\sqrt{-\Lambda}}-z_0} dz \ 2 e^{A(z)} 
(\partial_z \rho(z))^2. 
\label{23}
\end{eqnarray}
Here we have carefully kept the KK-mass term since we wish to examine
whether this solution leads to massless 'photon' on a brane. 
The localization condition of this mode on a brane requires
the first integral over $z$ to be finite. Thus let us focus on this
integral first. 
\begin{eqnarray}
I_1^{(1)} &{}& \equiv \int_0^{\frac{\pi}{\sqrt{-\Lambda}}-z_0} dz 
e^{A(z)} \rho^2(z) \nn\\
&{}& = \int_0^{\pi - \varepsilon} dw \frac{L}{\sin (w + \varepsilon)}
\left[ \frac{\alpha}{L \Lambda} \cos (w + \varepsilon) + \beta \right]^2. 
\label{24}
\end{eqnarray}
This integral $I_1^{(1)}$ is in general divergent, but only when
the equality $\frac{\alpha}{L \Lambda} = \beta$ holds, it becomes
to be finite. Henceforth, we shall consider this specific case.
Then, it is straightforward to calculate the above integral as well
as the second integral over $z$ in Eq. (\ref{23}) associated with
the KK-mass term whose result is given by
\begin{eqnarray}
S_1^{(0)} &=& -\frac{1}{4} \int d^4 x \sqrt{-\hat{g}} \hat{g}^{\mu\nu} 
\hat{g}^{\lambda\sigma} f_{\mu\lambda}f_{\nu\sigma}
( -\frac{2 \alpha^2}{L \Lambda^2}) \left[ \cos^2 \frac{\varepsilon}{2}
+ \log (\sin^2 \frac{\varepsilon}{2}) \right] \nn\\
&{}& -\frac{1}{4} \int d^4 x \sqrt{-\hat{g}} \hat{g}^{\mu\nu} 
a_\mu a_\nu \frac{4 \alpha^2}{L (-\Lambda)} \cos^2 \frac{\varepsilon}{2}. 
\label{25}
\end{eqnarray}
The quantities in front of the kinetic and the mass terms are obviously 
finite, so the gauge field is localized on an $AdS_4$ brane, which
is contrasted with the case of a Minkowski brane \cite{Oda3}.

At this stage, we can take a step further. Namely, provided that
when $\varepsilon \approx 0$ we redefine the brane gauge field 
$a_\mu$ as
\begin{eqnarray}
\sqrt{-\frac{2 \alpha^2}{L \Lambda^2} \left[ \cos^2 
\frac{\varepsilon}{2} + \log (\sin^2 \frac{\varepsilon}{2}) \right]} 
\ a_\mu \rightarrow a_\mu, 
\label{26}
\end{eqnarray}
Eq. (\ref{25}) reads
\begin{eqnarray}
S_1^{(0)} &=& -\frac{1}{4} \int d^4 x \sqrt{-\hat{g}} \left[
\hat{g}^{\mu\nu} \hat{g}^{\lambda\sigma} f_{\mu\lambda}f_{\nu\sigma}
+ 2 \Lambda \frac{\cos^2 \frac{\varepsilon}{2}} 
{\cos^2 \frac{\varepsilon}{2} + \log (\sin^2 \frac{\varepsilon}{2})}  
\hat{g}^{\mu\nu} a_\mu a_\nu \right].
\label{27}
\end{eqnarray}
This expression implies that the mass of the brane gauge field is
very tiny if the brane cosmological constant $|\Lambda| \approx 
\varepsilon^2$ is small enough. In the work of Karch and Randall
\cite{Karch} the small brane cosmological constant is needed to
make contact with experiment. In the present analysis the smallness
of the brane cosmological constant is obtained from the physical requirement 
that the $U(1)$ gauge field $a_\mu$ must be $\it{massless}$ 'photon' 
on an $AdS_4$ brane.
As a final remark in this paragraph, let us notice that owing to the equality 
$\frac{\alpha}{L \Lambda} = \beta$, our solution (\ref{22}) reduces
to the form
\begin{eqnarray}
\rho(z)  = - \frac{2 \alpha}{L (- \Lambda)} \cos^2 
\frac{\sqrt{- \Lambda}}{2}(z + z_0). 
\label{28}
\end{eqnarray}
Like the graviton as well as a real scalar, this solution satisfies
the box boundary condition at $z = \frac{\pi}{\sqrt{- \Lambda}}
-z_0$, where $\rho(z) = 0$. It is quite of interest that the requirement 
of the localization for the gauge field naturally leads to the same 
boundary condition as the other bosonic fields. As an important remark, 
we wish to point out the following fact about
the zero-mode wave function (28). This zero-mode function has the form
of cosine curve so that it spreads along the fifth dimension rather widely
compared with the zero-mode of the other fields with the exponentially
damping form.  Thus it might be not appropriate to use the word,
"localization", 
in the case of vector field. However, as shown above, the amplitude is
certainly 
peaked on the brane.  It should be understood from this context that we have
also used the word, "localization", even in this case.

Finally, let us consider the gravitino field of spin 3/2.
It is well known that even if free field actions for any higher spin 
do indeed exist, we cannot construct interacting actions for more than 
spin 2 at least within the framework of the local field theory. Thus, it
is now sufficient to consider the remaining spin 3/2 case for the
study of localization of the whole spin fields. (As mentioned before,
transverse traceless graviton modes of spin 2 have the same localization
property as a real scalar. And the higher-rank antisymmetric tensor
fields can be treated in a similar way to the electro-magnetic field.)
The trapping of the gravitino might be automatic when the graviton is 
trapped and the theory is supersymmetrized  since the gravitino is 
anyway a superpartner of the graviton, but at the present time of writing 
this article we do not have a grasp of a supersymmetric theory 
corresponding to Karch-Randall model, so it is valuable to pursue the 
issue along the same line of arguments as above.
 
The action for spin 3/2 bulk gravitino is given by the Rarita-Schwinger 
action \cite{Oda3}
\begin{eqnarray}
S_{3/2} = \int d^5 x \sqrt{-g} \bar{\Psi}_M i \Gamma^{[M} \Gamma^N
\Gamma^{R]} (D_N + \delta_N^z \Gamma_z m \varepsilon(z)) \Psi_R, 
\label{29}
\end{eqnarray}
where $D_M \Psi_N = \partial_M \Psi_N - \Gamma^R_{MN} \Psi_R + 
\frac{1}{4} \omega_M^{AB} \gamma_{AB} \Psi_N$ and the square bracket
denotes the anti-symmetrization with weight 1. From the metric
condition $D_M e_N^A = \partial_M e_N^A - \Gamma^R_{MN} e_R^A + 
\omega_M^{AB} e_{NB} = 0$, we obtain the concrete expression of the
affine connections  $\Gamma^R_{MN} = e^R_A (\partial_M e_N^A 
+ \omega_M^{AB} e_{NB})$. In the background (\ref{1}), the affine
connections are calculated to be 
\begin{eqnarray}
\Gamma^\rho_{\mu\nu} &=& \hat{\Gamma}^\rho_{\mu\nu}(\hat{e}), \
\Gamma^\rho_{\mu z} = \Gamma^\rho_{z \mu} = A'(z) \delta^\rho_\mu,
\nn\\
\Gamma^z_{\mu\nu} &=& - A'(z) \hat{g}_{\mu\nu}, \ \Gamma^z_{zz} 
= A'(z), \ otherwise = 0. 
\label{30}
\end{eqnarray}
With the gauge condition $\Psi_z = 0$, the nonvanishing components of
$D_M \Psi_N$ read
\begin{eqnarray}
D_\mu \Psi_\nu &=& \hat{D}_\mu(\hat{e}) \Psi_\nu + \frac{1}{2} A'(z)
\gamma_\mu \gamma_z \Psi_\nu, \nn\\
D_\mu \Psi_z &=& - A'(z) \Psi_\mu, \nn\\
D_z \Psi_\mu &=& (\partial_z - A'(z)) \Psi_\mu,
\label{31}
\end{eqnarray}
where we have used Eqs. (\ref{15}) and (\ref{30}). And we have defined
$\hat{D}_\mu(\hat{e}) \Psi_\nu
\equiv \partial_\mu \Psi_\nu - \hat{\Gamma}^\rho_{\mu\nu} \Psi_\rho + 
\hat{\omega}_\mu(\hat{e}) \Psi_\nu$.

The equations of motion $\Gamma^{[M} \Gamma^N \Gamma^{R]} (D_N + 
\delta_N^z \Gamma_z m \varepsilon(z)) \Psi_R = 0$ can be cast to the form 
\begin{eqnarray}
g^{\mu\nu} \left[\Gamma^z (\partial_z + A'(z)) + m \varepsilon(z)
\right] \Psi_\nu = 0, 
\label{32}
\end{eqnarray}
where we have used equations $\gamma^\mu \Psi_\mu = \hat{D}^\mu \Psi_\mu
= \gamma^{[\mu} \gamma^\nu \gamma^{\rho]} \hat{D}_\nu \Psi_\rho = 0$.
Let us look for a solution with the form $\Psi_\mu(x^M) = \psi_\mu 
(x^\lambda) v(z)$. If the chirality condition $\Gamma^z \psi_\mu = 
\psi_\mu$ is utilized in Eq. (\ref{32}), we can get a solution 
$v(z) = v_0 e^{-A(z) - m \varepsilon(z) z}$ in a perfectly analogous
manner to the case of spin 1/2 spinor.

Substituting this solution into the action (\ref{29}), we arrive at 
the following expression
\begin{eqnarray}
S_{3/2}^{(0)} &=& \int d^5 x \sqrt{-g} \bar{\Psi}_M^{(0)}
i \Gamma^{[M} \Gamma^N \Gamma^{R]} (D_N + \delta_N^z \Gamma_z m 
\varepsilon(z)) \Psi_R^{(0)} \nn\\
&=& \int d^4 x \sqrt{-\hat{g}} \bar{\psi}_\mu i \gamma^{[\mu} 
\gamma^\nu \gamma^{\rho]} \hat{D}_\nu \psi_\rho
\int_0^{\frac{\pi}{\sqrt{-\Lambda}}-z_0} dz e^{2A(z)} v^\dagger(z) 
v(z) + \cdots. 
\label{33}
\end{eqnarray}
Again the condition for the localization of the gravitino on a
brane requires the integral over $z$ to take a finite value.
Indeed this statement can be checked as follows:
\begin{eqnarray}
I_{3/2} &{}& \equiv \int_0^{\frac{\pi}{\sqrt{-\Lambda}}-z_0} dz 
e^{2A(z)} v^\dagger(z) v(z) \nn\\
&{}& = |v_0|^2 \int_0^{\frac{\pi}{\sqrt{-\Lambda}}-z_0} dz 
e^{- 2 m z} < \infty.
\label{34}
\end{eqnarray}
At this stage, it is of interest to notice that the condition has
the same form as in spin 1/2 spinor field up to an irrelevant constant, 
so whenever spin 1/2 fermion is localized, spin 3/2 gravitino is also 
localized on a brane.

In conclusion, in this article we have presented a complete
analysis of localization of all bulk fields on an $AdS_4$ brane
in $AdS_5$ from the viewpoint of the local field theory. 
We can summarize the results obtained in this article as follows. 
It has been shown that the whole bulk fields can be localized
on an $AdS_4$ brane by only the gravitational interaction without
invoking additional interactions.
In particular, for the localization of fermionic fields we do not have to 
introduce a mass term with a 'kink' profile into the action. 
Moreover, the existence of the massless gauge field on the 
brane requires that the brane cosmological constant should be small
enough not to contradict the standard model. 

Accordingly, it seems that Karch-Randall model \cite{Karch} gives us 
an interesting phenomenological model. A direction for future study is 
to examine whether this model could also provide a realistic
cosmological model \cite{Oda5}. Another interesting direction is
to supersymmetrize the model at hand.

\vs 1


\end{document}